\documentclass[aps,prb,reprint,groupedaddress,showpacs]{revtex4-1}

\usepackage{graphicx,amsmath}

\begin{document}

\title{Disorder-induced double resonant Raman process in graphene}

\author{J. F. Rodriguez-Nieva,$^{1}$ E. B. Barros,$^{1,2}$ R. Saito,$^{3}$ and M. S. Dresselhaus$^{1,4}$}
\affiliation{$^{1}$ Department of Physics, Massachusetts Institute of Technology, Cambridge, MA 02139, USA}
\affiliation{$^{2}$ Departamento de F\'{i}sica, Universidade Federal do Cear\'{a}, Fortaleza, Cear\'{a} 60455-760, Brazil}
\affiliation{$^{3}$ Department of Physics, Tohoku University, Sendai 980-8578, Japan}
\affiliation{$^{4}$ Department of Electrical Engineering and Computer Science, Massachusetts Institute of Technology, Cambridge, MA 02139, USA}


\begin{abstract}

An analytical study is presented of the double resonant Raman scattering process in graphene, responsible for the D and D$^{\prime}$ features in the Raman spectra. This work yields analytical expressions for the D and D$^{\prime}$ integrated Raman intensities that explicitly show the dependencies on laser energy, defect concentration, and electronic lifetime. Good agreement is obtained between the analytical results and experimental measurements on samples with increasing defect concentrations and at various laser excitation energies. The use of Raman spectroscopy to identify the nature of defects is discussed. Comparison between the models for the edge-induced and the disorder-induced D band intensity suggests that edges or grain boundaries can be distinguished from disorder by the different dependence of their Raman intensity on laser excitation energy. Similarly, the type of disorder can potentially be identified not only by the intensity ratio $I_{\mathrm{D}}/I_{\mathrm{D}^{\prime}}$, but also by its laser energy dependence. Also discussed is a quantitative analysis of quantum interference effects of the graphene wavefunctions, which determine the most important phonon wavevectors and scattering processes responsible for the D and D$^{\prime}$ bands.

\end{abstract}

\pacs{78.30.-j,78.67.Wj,81.05.ue}
\keywords{}

\maketitle

\section{Introduction}

Raman spectroscopy is a powerful non-destructive characterization technique that provides invaluable information about graphitic samples,\cite{milliebook,saitoreview,ferrarireview} such as phonon properties,\cite{phonondisp,ferrarireview2,malardreview} doping,\cite{doping1,doping2} and the number of layers\cite{gphlayers2006} for both few-layer graphenes and carbon nanotubes. In particular, the D and D$^{\prime}$ bands ($\sim$1350 cm$^{-1}$ and $\sim$1620 cm$^{-1}$ for 2.4-eV laser excitation energy, $E_{\mathrm{L}}$, respectively) originate from the presence of defects in the sample, such as grain boundaries\cite{edges1,edges2,edges3} or point defects.\cite{pimentadefect,cancadodefect} For this reason, these defect-induced Raman features, distinct from the defect-free G band ($\sim$1585 cm$^{-1}$) and the G$^{\prime}$ band ($\sim$2680 cm$^{-1}$), have been widely used to assess the graphene materials' quality when used in graphene-based devices.\cite{ferrarireview2}

The origin of the D and D$^{\prime}$ bands has been previously discussed by several authors by using the characteristics of the so-called double resonant (DR) Raman scattering process.\cite{drthomsen,c887,f1020,thomsen2004,reichreview,pimentadefect,ramanreview}  This explanation has been successfully applied to qualitatively describe some of the important aspects of the D and D$^{\prime}$ bands. Most notably, the dispersive behavior of the D-band Raman shift\cite{dispersive1,dispersive2} as a function of $E_{\mathrm{L}}$ was successfully explained within the DR picture.

Despite the numerous theoretical and experimental works on the DR process, some of the most interesting and potentially useful questions about the characterization of defects in graphene remain to be answered. For instance, the distinguishing signatures of the different types of defects regarding the Raman spectra remain an open problem. Do edges or grain boundaries have different fingerprints in the Raman spectra than those for point defects? Do all defects have the same laser energy dependence? Are the D and D$^{\prime}$ bands affected differently by each type of point defect? Ultimately, the open question that needs to be addressed is whether Raman spectroscopy can be used as an accurate and non-destructive tool to, not only quantify, but also to distinguish and characterize specific defects from one another in $sp^2$ graphitic materials.

In this paper, we present a detailed analytical study that describes the integrated D and D$^{\prime}$ Raman intensities in order to address the above-mentioned questions. Our results provide new insights about the Raman physics in graphene which were previously overlooked, and complements more detailed numerical calculations.

\begin{figure}[t]
\centering \includegraphics[scale=1.0]{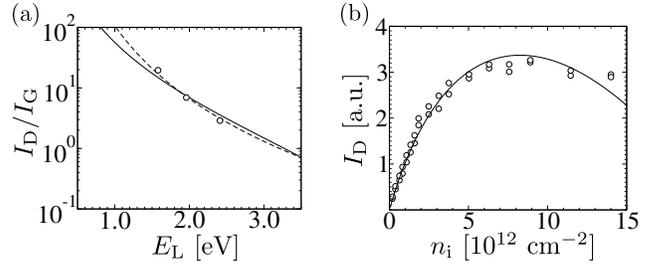}
\caption{(a) Laser energy dependence of the integrated Raman intensity ratio $I_{\mathrm{D}}/I_{\mathrm{G}}$ between the D and G bands obtained from Eq. (\ref{eq:Idr}) (solid line), and experimental points from Ref. \onlinecite{cancadodefect}. For the $I_{\mathrm{G}}$ intensity, we use the standard textbook dependence of $I_{\mathrm{G}}\propto E_{\mathrm{L}}^4$.\cite{basko2008} The dashed line indicates the frequently used $I_{\mathrm{D}}/I_{\mathrm{G}}\propto E_{\mathrm{L}}^{-4}$ fit. (b) The integrated D-band intensity as a function of defect concentration $n_{\mathrm{i}}$ obtained from Eq. (\ref{eq:Idr}) (solid line), and experimental points of Ref. \onlinecite{concentration2}. }
\label{fig:exp}
\end{figure}

Several experimental results have already paved the way for progress in understanding the DR physics. For example, the laser energy $E_{\mathrm{L}}$ dependence of the frequently used $I_{\mathrm{D}}/I_{\mathrm{G}}$ ratio between the D-band and the G-band intensities has been measured by many groups on samples with various types of defects\cite{defectreview} and at different concentrations, thereby providing a large body of information about defects. While some samples\cite{pimentadefect,cancadodefect} show an $I_{\mathrm{D}}/I_{\mathrm{G}}\propto E_{\mathrm{L}}^{-4}$ dependency [see Fig. \ref{fig:exp}(a)], other measurements have shown a smaller power-law exponent.\cite{barros2007,eckmann2013} Furthermore, it was recently shown by Eckmann \textit{et al}.\cite{eckmann2013} that, even within a single sample, the Raman intensities of the D and D$^{\prime}$ bands can have different laser energy dependencies, as well as suggesting that the D and D$^{\prime}$ intensity ratio can be different depending on the type of defect.\cite{eckmann2012} Since the D and D$^{\prime}$ bands originate from, respectively, intervalley and intravalley elastic scattering of the photoexcited electron-hole pair, the scattering potential should play an important role in determining the Raman scattering amplitude. 

In addition, several studies have focused on the dependence of the integrated D-band intensity as a function of defect concentration.\cite{concentration,concentration2} In its simplest approximation, the integrated intensity depends linearly on the defect concentration. However, experimental measurements show that $I_{\mathrm{D}}$ reaches a peak value at a sufficiently large concentration of defects [see Fig. \ref{fig:exp}(b)], when the average distance $L_{\mathrm{d}}$ between defects is $\sim$3 nm.\cite{concentration}

Numerical calculations of the Raman cross section have previously been the dominant procedure used to model the features of the Raman spectra induced by several types of defects. In this way, several authors studied the problem of disorder,\cite{venezuela2011} edges,\cite{barros2011} grain boundaries,\cite{Sato2006} and isotope impurities.\cite{isotopes} Given that the DR process is a fourth-order process involving interactions between electrons, phonons, photons, and defects, and requires knowledge of the phonon dispersion relations, electronic band structure, and electron lifetimes, numerical techniques provide a powerful and effective way to address the defect problem. However, the above-mentioned experimental observations are difficult to understand directly from calculations.

Alternatively, analytic calculations require a series of approximations which affect the predictive power of the resulting model, but allow for a more insightful analysis into the underlying physics involved. One notable step in this direction was taken by Basko.\cite{basko2007,basko2008,baskoedge} There, the author obtained analytical expressions for the Raman intensity for the G$^{\prime}$ band\cite{basko2007,basko2008} and for the edge-induced D band.\cite{baskoedge} For instance, power-law dependencies on the inverse electron lifetime $\gamma$ of the integrated Raman intensity of the G$^{\prime}$ band and its overtones were obtained, suggesting the use of the ratio of these Raman intensities to indirectly measure the pertinent electronic lifetimes.\cite{basko2007}

Interestingly, both edges and disorder produce a D-band feature in the Raman spectra. However, the description of the intermediate states in the edge-induced Raman scattering case\cite{baskoedge} already incorporates eigenstates in the presence of the edge (i.e., scattered states instead of plane waves), while the DR picture used to describe the disorder-induced Raman scattering uses plane waves perturbed by an external potential. Therefore, the edge-induced Raman scattering is studied as a third-order process,\cite{baskoedge} while the disorder-induced Raman scattering is studied as a fourth-order process.\cite{drthomsen,c887} Then, a comparison between the predictions for the D band induced by these two types of defects is necessary.

In this work, we do a detailed analytical study of the DR theory which brings to light the role played by the different parameters of the model, such as the laser energy, scattering potential, and electronic lifetimes. For this purpose, we obtain analytic expressions for the disorder-induced Raman intensity within the DR theory using an effective Hamiltonian description. We do a comparison between our model and recent experimental measurements, and discuss the main features of our results in relation to the above-mentioned experimental observations. Furthermore, we compare our results with the analytical models obtained for the edge-induced D band.\cite{baskoedge} Our analysis yields, additionally, a quantitative discussion of phase interference effects.\cite{thomsen2004,venezuela2011}

The outline of the paper is as follows: In Sec. IIA we briefly review the theory of the DR Raman process, and in Sec. IIB we describe the relevant matrix elements. In Sec. III we make a detailed analysis of the DR Raman intensity, quantifying the contribution from each of several different scattering processes that are possible, and the main results are discussed in Sec. IV. The conclusions are given in Sec. V.

\section{Theory}

\subsection{Raman intensity calculation}
\label{sec:theorya}

\begin{figure}
\centering \includegraphics[scale=1.0]{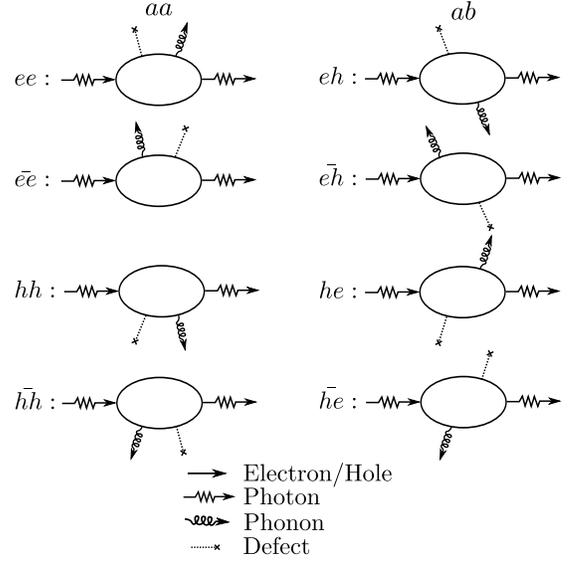}
\caption{Diagrams contributing to the double resonant Raman scattering process. The notation $ab$ ($\bar{ab}$) indicates that the particle $a$ ($a=e,h$) is scattered first by a defect (phonon), and particle $b$ ($b=e,h$) is scattered next by a phonon (defect), and where $e$ ($h$) stands for electron (hole).}
\label{fig:diagrams}
\end{figure}

The DR process is understood as an inelastic fourth-order process that involves interactions of photoexcited electron-hole pairs with phonons and defects. Referring to Fig. \ref{fig:diagrams} and neglecting finite-temperature effects, we consider only Stokes scattering. The photoabsorption in its initial state is described by an incoming photon with momentum $\mathbf{Q}_{\mathrm{i}}$, energy $E_{\mathrm{L}}$, and polarization $\lambda_{\mathrm{i}}$, and the graphene system (electrons and phonons) is initially in its ground state. The possible final states are described by the production of a phonon with momentum $\mathbf{q}_{\mathrm{ph}}$, mode $\alpha$, and frequency $\omega_{\mathbf{q}_{\mathrm{ph}},\alpha}$, a photon with momentum $\mathbf{Q}_{\mathrm{f}}$ and polarization $\lambda_{\mathrm{f}}$, and the graphene electronic system is back to its ground state. Elastic scattering with a defect is necessary in order to guarantee momentum conservation in the DR process. 

In this paper, we compute the DR Raman scattering probability $\mathcal{I}_{\mathrm{DR}}$, defined as the total DR Raman probability of an incoming photon with momentum $\mathbf{Q}_{\mathrm{i}}$ and polarization $\lambda_{\mathrm{i}}$. The electromagnetic field is assumed to be confined in a box of volume $V=A L_z$, where $A$ is the area of the graphene layer and $L_{z}$ is the length of the box in the direction normal to the graphene plane. Then, $\mathcal{I}_{\mathrm{DR}}$ is calculated ($\hbar =1$) as
\begin{equation}
\mathcal{I}_{\mathrm{DR}}=\frac{2\pi L_z}{c}\sum_{\substack{\mathbf{Q}_{\mathrm{f}},\lambda_{\mathrm{f}}\\ \mathbf{q}_{\mathrm{ph}},\alpha}} |\mathcal{M}(\mathbf{q}_{\mathrm{ph}},\alpha)|^{2} \delta(E_{\mathrm{L}}-c|\mathbf{Q}_{\mathrm{f}}| - \omega_{\mathbf{q}_{\mathrm{ph}},\alpha}),
\label{eq:idr}
\end{equation}
where $c$ is the speed of light, $E_{\mathrm{L}}=c|\mathbf{Q}_{\mathrm{i}}|$, and the matrix $\mathcal{M}(\mathbf{q}_{\mathrm{ph}},\alpha)=\sum_p\mathcal{M}_p(\mathbf{q}_{\mathrm{ph}},\alpha)$ describing the Raman scattering arises from consideration of all possible diagrams $p$ for the interactions, shown in Fig. \ref{fig:diagrams}. The Raman intensity $I_{\mathrm{DR}}$, which is the magnitude measured in experiments, is related to $\mathcal{I}_{\mathrm{DR}}$ by the simple relation $I_{\mathrm{DR}}=I_0 \times \mathcal{I}_{\mathrm{DR}}$, where $I_0$ is the intensity of incoming photons.

Following the notation introduced by Venezuela \textit{et al},\cite{venezuela2011} we label the $aa$ processes as those in which either only electrons or holes participate in the scattering (left column in Fig. \ref{fig:diagrams}), while $ab$ processes are those in which both electrons and holes participate in the scattering (right column in Fig. \ref{fig:diagrams}). Furthermore, we indicate in Fig. \ref{fig:diagrams} the notation used individually for each process $p$.

We focus mostly on the calculation of the matrix $\mathcal{M}_{p}(\mathbf{q}_{\mathrm{ph}},\alpha)$ for the $ee$ and $eh$ processes throughout this paper, given that extension to the remaining processes is straightforward. Explicitly, the matrices $\mathcal{M}_{ee}(\mathbf{q}_{\mathrm{ph}},\alpha)$ and $\mathcal{M}_{eh}(\mathbf{q}_{\mathrm{ph}},\alpha)$ for the diagrams $ee$ and $eh$ in Fig.\ref{fig:diagrams}, respectively, are given by 
\begin{widetext}
\begin{equation}
\mathcal{M}_{ee}({\mathbf{q}_{\mathrm{ph}},\alpha})=\sum_{\mathbf{p}\in\mathrm{BZ}}\frac{ \langle \mathbf{p},\pi | \mathcal{H}_{\mathrm{eR},\mathrm{f}}|\mathbf{p},\pi^* \rangle \langle \mathbf{p},\pi^* |\mathcal{H}_{\mathrm{ep},\alpha}| \mathbf{p+q}_{\mathrm{ph}},\pi^* \rangle \langle \mathbf{p+q}_{\mathrm{ph}},\pi^*| \mathcal{H}_{\mathrm{ed}} | \mathbf{p},\pi^* \rangle \langle \mathbf{p},\pi^*| \mathcal{H}_{\mathrm{eR},\mathrm{i}} | \mathbf{p},\pi \rangle }{(E_{\mathrm{L}}-\omega_{\mathbf{q}_{\mathrm{ph}},\alpha}-\varepsilon_{\mathbf{p}}^{\pi^*}+\varepsilon_{\mathbf{p}}^{\pi}-i\gamma/2)(E_{\mathrm{L}}-\varepsilon_{\mathbf{p+q}_{\mathrm{ph}}}^{\pi^*}+\varepsilon_{\mathbf{p}}^{\pi}-i\gamma/2)(E_{\mathrm{L}}-\varepsilon_{\mathbf{p}}^{\pi^*}+\varepsilon_{\mathbf{p}}^{\pi}-i\gamma/2)},
\label{eq:kee}
\end{equation}
and
\begin{equation}
\mathcal{M}_{eh}({\mathbf{q}_{\mathrm{ph}},\alpha})=-\sum_{\mathbf{p}\in \mathrm{BZ}}\frac{\langle \mathbf{p+q}_{\mathrm{ph}},\pi | \mathcal{H}_{\mathrm{eR},\mathrm{f}}|\mathbf{p+q}_{\mathrm{ph}},\pi^* \rangle \langle \mathbf{p},\pi |\mathcal{H}_{\mathrm{ep},\alpha}| \mathbf{p+q}_{\mathrm{ph}},\pi \rangle \langle \mathbf{p+q}_{\mathrm{ph}},\pi^*| \mathcal{H}_{\mathrm{ed}} | \mathbf{p},\pi^* \rangle \langle \mathbf{p},\pi^*| \mathcal{H}_{\mathrm{eR},\mathrm{i}} | \mathbf{p},\pi \rangle}{(E_{\mathrm{L}}-\omega_{\mathbf{q}_{\mathrm{ph}},\alpha}-\varepsilon_{\mathbf{p+q}_{\mathrm{ph}}}^{\pi^*}+\varepsilon_{\mathbf{p+q}_{\mathrm{ph}}}^{\pi}-i\gamma/2)(E_{\mathrm{L}}-\varepsilon_{\mathbf{p+q}_{\mathrm{ph}}}^{\pi*}+\varepsilon_{\mathbf{p}}^{\pi}-i\gamma/2)(E_{\mathrm{L}}-\varepsilon_{\mathbf{p}}^{\pi^*}+\varepsilon_{\mathbf{p}}^{\pi}-i\gamma/2)},
\label{eq:keh}
\end{equation}
\end{widetext}
where the summation in electronic momentum $\mathbf{p}$ is taken over the graphene hexagonal Brillouin zone (BZ), $\mathcal{H}_{\mathrm{eR}}$, $\mathcal{H}_{\mathrm{ep}}$, and $\mathcal{H}_{\mathrm{ed}}$ denote the electron-radiation, electron-phonon, and electron-defect interactions, respectively,  $\pi$($\pi^*$) denotes the hole (electron) band, $\varepsilon_{\mathbf{p}}^\pi$ ($\varepsilon_{\mathbf{p}}^{\pi^*}$) is the energy of a hole (electron) with wave vector $\mathbf{p}$, and $\gamma$ is the electronic broadening. In particular, we assume that $\gamma=\gamma_{\mathrm{ep}}+\gamma_{\mathrm{ed}}$ has contributions from electron-phonon scattering ($\gamma_{\mathrm{ep}}\sim$ meV) or electron-defect scattering ($\gamma_{\mathrm{ed}}\sim$ meV), and that, in comparison, the contribution from electron-photon scattering ($\gamma_{\mathrm{eR}}\sim\mu$eV) can be neglected. At electronic energies comparable to those of photons in the visible range, a value of $\gamma_{\mathrm{ep}}\sim15$ meV is obtained.\cite{elifetime} The value of $\gamma_{\mathrm{ed}}$ can be calculated from Fermi's golden rule $\gamma_{\mathrm{ed}}=2\pi\sum_{\mathbf{p}} |\langle \mathbf{p'}| \mathcal{H}_{\mathrm{ed}} | \mathbf{p} \rangle|^2\delta(\varepsilon_{\mathbf{p}}-\varepsilon_{\mathbf{p}'})$, where $\varepsilon_{\mathbf{p}}\sim E_{\mathrm{L}}/2$ (see Sec. \ref{sec:exp} for details). Furthermore, we consider throughout this work that $\gamma(\sim 10 \mathrm{meV}) \ll \omega_{\mathbf{q}_{\mathrm{ph}},\alpha}(\sim 0.2 \mathrm{eV}) \ll E_{\mathrm{L}}(\sim 2\mathrm{eV})$, which is the typical situation in experiments.

The characteristic feature of the DR process is that two of the three  denominators  in Eqs. (\ref{eq:kee}) and (\ref{eq:keh}) can be simultaneously zero at specific points in phonon and electronic phase space, and thus the name double resonance.\cite{drthomsen} This is different than the G$^{\prime}$ band case (two-phonon scattering around 2700 cm$^{-1}$), where a triple resonance is possible.\cite{basko2007}

Raman measurements yield the number of outgoing photons coming to a detector covering a solid angle $\Omega_{\mathrm{f}}$. In order to make direct comparison with experiments, we express $\mathcal{I}_{\mathrm{DR}}$ in Eq. (\ref{eq:idr}) per unit solid angle $\Omega_{\mathrm{f}}$. The summation over outgoing photon momentum $\mathbf{Q}_{\mathrm{f}}$ can be written as an integral in spherical coordinates given by $\sum_{\mathbf{Q}_{\mathrm{f}}}=(V/8\pi^3)\int d\mathrm{Q}_{\mathrm{f}} \int d\Omega_{\mathrm{f}} \mathrm{Q}_{\mathrm{f}}^2$, where $d\Omega_{\mathrm{f}}$ is the differential solid angle covered by the outgoing photons. In Eq. (\ref{eq:idr}), the matrix $\mathcal{M}$ only depends on the direction $\hat{\mathbf{Q}}_{\mathrm{f}}$ and polarization $\lambda_{\mathrm{f}}$ of the outgoing photon, but not on $|\mathbf{Q}_{\mathrm{f}}|$, given its small value. Then, energy conservation dictates $c|\mathbf{Q}_{\mathrm{f}}|=E_{\mathrm{L}}-\omega_{\mathbf{q}_{\mathrm{ph}},\alpha}$, and the delta function in Eq. (\ref{eq:idr}) is absorbed upon integration on $d \mathrm{Q}_{\mathrm{f}}$. Therefore, we obtain
\begin{equation}
\frac{d\mathcal{I}_{\mathrm{DR}}}{d\Omega_{\mathrm{f}}}=\frac{VL_zE_{\mathrm{L}}^2}{4\pi^2c^4}\sum_{\mathbf{q}_{\mathrm{ph}} , \alpha , \lambda_{\mathrm{f}} }|\sum_{p}\mathcal{M}_{p}(\mathbf{q}_{\mathrm{ph}},\alpha)|^2,
\label{eq:idr2}
\end{equation}
where we used $c|\mathbf{Q}_{\mathrm{f}}|\approx E_{\mathrm{L}}$. The values of $\mathcal{M}_{p}$ obtained from the diagrams in Fig. \ref{fig:diagrams} can be used as input for Eq. (\ref{eq:idr2}) to obtain $d\mathcal{I}_{\mathrm{DR}}/d\Omega_{\mathrm{f}}$. In our calculations below, we assume unpolarized and normally incident photons, and the detection of backscattered photons in all polarization directions. Furthermore, because the LO and $A_1$ Raman-active modes produce a Raman shift much larger than their respective linewidth, we can separate in Eq. (\ref{eq:idr2}) the contribution from each of these modes to the integrated Raman intensity.

\subsection{Effective Hamiltonian Description}
\label{sec:theoryb}

In the long-wavelength limit, the electronic states in the vicinity of the $K$ and $K^{\prime}$ points in the BZ, with momenta $\mathbf{p}=\mathbf{K}+\mathbf{k}$ and $\mathbf{p}=\mathbf{K}^{\prime}+\mathbf{k}$, respectively, and $\mathbf{k}$ a small wave vector relative to the BZ scale, can be described by the massless Dirac Hamiltonian
\begin{equation}
\mathcal{H}_{0}=v_{\mathrm{F}} \int d\mathbf{r} \, \psi^{\dagger}(\mathbf{r})  \left( \begin{array}{cc} \boldsymbol\sigma \cdot \hat{\mathbf{k}} & 0 \\ 0 & \boldsymbol\sigma^* \cdot \hat{\mathbf{k}} \end{array} \right) \psi(\mathbf{r}),
\label{eq:H0}
\end{equation}
where $\psi(\mathbf{r})$ is the four-component spinor describing electrons in the two graphene sublattices and in each of the two valleys, $\hat{\mathbf{k}}=-i\nabla_{\mathbf{r}}$, $v_{\mathrm{F}}$ is the Fermi velocity, and $\boldsymbol \sigma = (\sigma_x,\sigma_y)$ are Pauli matrices. Because in this description the wavefunctions acquire a new pseudospin index $s$ that labels the valley $s=K, K^{\prime}$, then it is necessary to replace the summation subindex in Eqs. (\ref{eq:kee}) and (\ref{eq:keh}) as $\sum_{\mathbf{p}\in \mathrm{BZ}}\rightarrow \sum_{\mathbf{k}s}$. Furthermore, it is important to note that, within the effective Hamiltonian approximation, intervalley transitions are described in terms of a change in the valley spin-index.

The electron-photon coupling can be obtained by the requirement of gauge invariance $\hat{\mathbf{k}}\rightarrow \hat{\mathbf{k}}-(e/c)\mathbf{A}$ in Eq. (\ref{eq:H0}), where $\mathbf{A}$ is the vector potential.\cite{sasaki2008} Then, $\mathcal{H}_{\mathrm{eR}}$ is given by
\begin{equation}
\mathcal{H}_{\mathrm{eR}}=-\frac{ev_{\mathrm{F}}}{c}\int d\mathbf{r} \, \psi^{\dagger}(\mathbf{r}) \left( \begin{array}{cc}  \boldsymbol{\sigma}\cdot \mathbf{A}(\mathbf{r}) & 0 \\ 0 & \boldsymbol{\sigma}^* \cdot \mathbf{A}(\mathbf{r}) \end{array} \right) \psi(\mathbf{r}),
\end{equation}
where $\mathbf{A}(\mathbf{r})$ is 
\begin{equation}
\mathbf{A}(\mathbf{r})=\sum_{\mathbf{Q}\lambda} \sqrt{\frac{2\pi c}{V|\mathbf{Q}|}}\left( a_{\mathbf{Q}\lambda}\mathbf{e}_{\mathbf{Q}\lambda}+a_{-\mathbf{Q}\lambda}^{\dagger}\mathbf{e}_{-\mathbf{Q}\lambda}^{*} \right)e^{i\mathbf{Q}\cdot \mathbf{r}}.
\end{equation}

The electron-phonon interaction can be modeled by considering the variation in the tight-binding hopping parameter induced by the change in the carbon-carbon bond length due to lattice vibrations. Given that we are interested in zone-center and zone-boundary phonons, we denote $\mathbf{q}_{\mathrm{ph}}=\mathbf{q}_{\mu}+\mathbf{q}$ ($\mu=\Gamma,K$), where $\mathbf{q}_{\Gamma}=0$ is the $\Gamma$ point in the graphene BZ, $\mathbf{q}_{K}=\mathbf{K}$ is the $K$ point in the graphene BZ, and $\mathbf{q}$ a small wave vector relative to the BZ scale. Furthermore, for the DR Raman process we only need to include the zone-center LO phonon mode (responsible for the D$^{\prime}$ band), and the zone-boundary $A_1$ phonon mode (responsible for the D band), which are the Raman active modes. Thus, for compactness, $\mu=\Gamma$ hereafter refers particularly to the LO mode in the vicinity of the $\Gamma$ point, while $\mu=K$ refers to the $A_1$ mode in the vicinity of the $K$ point.

The electron-phonon coupling term $\mathcal{H}_{\mathrm{ep}}$ for both zone-center\cite{andointra,sasaki2008} and zone-boundary\cite{andointer} phonons is then given by
\begin{equation}
\begin{array}{cc} \mathcal{H}_{\mathrm{ep}}=  & \displaystyle -i \int d\mathbf{r}\psi^{\dagger} (\mathbf{r}) \left[ F_{\Gamma} \left( \begin{array}{cc} \boldsymbol\sigma \times \mathbf{u}(\mathbf{r}) & 0 \\ 0 & -\boldsymbol\sigma^* \times \mathbf{u}(\mathbf{r}) \end{array} \right)  \right. \\ & \\ & \left. -F_{K} \left( \begin{array}{cc} 0 & w^{*}(\mathbf{r})\sigma_{y} \\ w(\mathbf{r})\sigma_{y} & 0 \end{array} \right) \right] \psi(\mathbf{r}), \end{array}
\label{eq:hep}
\end{equation}
where the parameters $F_{K}$ and $F_{\Gamma}$ ($F_{\Gamma}=F_{K}/\sqrt{2}$) are the force constants for intervalley and intravalley scattering, respectively. In Eq. (\ref{eq:hep}), the zone-center displacement field $\mathbf{u}(\mathbf{r})$ caused by the LO phonon mode with frequency $\omega_{\mathbf{q},\Gamma} = \omega_{\mathbf{q},\mathrm{LO}}$ is given by
\begin{equation}
\mathbf{u}(\mathbf{r})=\sum_{\mathbf{q}} \sqrt{\frac{1}{A\rho\,\omega_{\mathbf{q},\Gamma}}} \left( b_{\mathbf{q},\mathrm{LO}}\mathbf{e}_{\mathbf{q}}+b_{-\mathbf{q},\mathrm{LO}}^{\dagger}\mathbf{e}_{-\mathbf{q}}\right)e^{i\mathbf{q}\cdot \mathbf{r}},
\label{eq:epu}
\end{equation}
where $\rho$ is the mass density of graphene, and $\mathbf{e}_{\mathbf{q}}=(q_{x},q_{y})/|\mathbf{q}|$ is the LO polarization vector of the phonon amplitude. Alternatively, the zone-boundary distortion $w(\mathbf{r})$ induced by the $A_{1}$ phonon mode\cite{andointer} with frequency $\omega_{\mathbf{q},K}= \omega_{\mathbf{K+q},A_1}$ is given by
\begin{equation}
w(\mathbf{r})=\sum_{\mathbf{q}} \sqrt{\frac{1}{A\rho\omega_{\mathbf{q},K}}} \left( b_{\mathbf{q},K}+b_{-\mathbf{q},K^{\prime}}^{\dagger}\right)e^{i\mathbf{q}\cdot \mathbf{r}},
\label{eq:epw}
\end{equation}
and couples eigenstates from valley $K$ with eigenstates of valley $K^{\prime}$.

For the electron-defect interaction, we consider defect potentials randomly distributed over the lattice at positions $\mathbf{r}_j$. Then, $\mathcal{H}_{\mathrm{ed}}$ is given by
\begin{equation}
\mathcal{H}_{\mathrm{ed}}=\int d\mathbf{r} \psi^{\dagger}(\mathbf{r}) \left[ \frac{1}{A}\sum_{j,\mathbf{q}}U_{\mathbf{q}}e^{i\mathbf{q}\cdot(\mathbf{r}-\mathbf{r}_j)}\right] \psi(\mathbf{r}),
\label{eq:Hd}
\end{equation}
where the $4\times 4$ matrix $U_{\mathbf{q}}$ has components
\begin{equation}
U_{\mathbf{q}}=\left(\begin{array}{cc} U_{\mathbf{q},\Gamma} & U_{\mathbf{q},K} \\ U_{\mathbf{q},K}^{\dagger} & U_{\mathbf{q},\Gamma} \end{array}\right).
\label{eq:Hdc}
\end{equation}
The $2\times 2$ matrices $U_{\mathbf{q},\mu}$ ($\mu=\Gamma,K$) in Eq. (\ref{eq:Hd}) are the Fourier components of the defect potential for the different sublattice degrees of freedom, and for intravalley ($\mu=\Gamma$) and intervalley ($\mu=K$) scattering. In general, the matrices $U_{\mathbf{q},\mu}$ may contain contributions from both on-site and hopping terms. For instance, in Ref. \onlinecite{andobackscatt1}, $U_{\mathbf{q},\mu}$ is calculated for onsite potentials. Given that the wave vector $\mathbf{q}$ probed by Raman spectroscopy varies with photon energy, it is important to take into consideration (at least in principle) the general wave vector dependence of $U_{\mathbf{q},\mu}$ in Eq. (\ref{eq:Hd}). Throughout the analysis we assume a general function $U_{\mathbf{q},\mu}$, but we will consider point-like defects when explicitly comparing with experiments in this work.

\section{Results}
\label{sec:results}

\subsection{Phase interference effects: Phonon momentum selectivity and relevant diagrams}

\begin{figure}
\centering \includegraphics[scale=1.0]{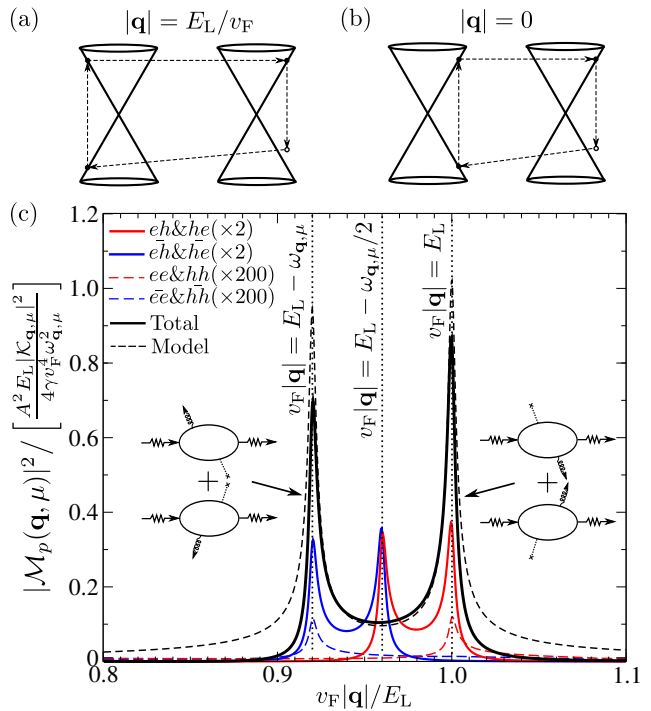}
\caption{Because of phase interference effects, only a small region of phonon phase space and a small number of diagrams in Fig. \ref{fig:diagrams} contribute dominantly to the Raman probability. For instance, (a) back-scattering of the photoexcited electron-hole pair by a phonon with momentum $\mathbf{q}_{\mathrm{ph}}=\mathbf{K}+\mathbf{q}$, where $|\mathbf{q}|=E_{\mathrm{L}}/v_{\mathrm{F}}$, provides a significantly larger contribution to the D-band Raman intensity than (b) forward scattering with $|\mathbf{q}|=0$.\cite{thomsen2004} (c) The contribution to the Raman matrix element $|\mathcal{M}|^2=|\sum_p \mathcal{M}_p|^2$ (black lines) is mainly due to $ab$ diagrams (colored solid lines) shown in the right column of Fig. \ref{fig:diagrams}.\cite{venezuela2011} On the contrary, $aa$ diagrams (colored dashed lines) have matrix elements $|\mathcal{M}_{p}|^2$ smaller than those of $ab$ processes by a factor of $(\omega_{\mathbf{q},\mu}/2E_{\mathrm{L}})^2$. At $v_{\mathrm{F}}|\mathbf{q}|=E_{\mathrm{L}}$, for example, both the $eh$ and $he$ diagrams provide the dominant contribution and, thus, $|\sum_{p}\mathcal{M}_{p}|^2$ is approximately four times the value of $\mathcal{M}_{eh}$. Note also the cancellation of the peaks at $v_{\mathrm{F}}|\mathbf{q}|=E_{\mathrm{L}}-\omega_{\mathbf{q},\mu}/2$, where all four $ab$ diagrams interfere destructively. The black dashed curve is obtained within our model from Eq. (\ref{eq:keh1}), valid only in the vicinity of each peak.}
\label{fig:schematics}
\end{figure}

Although the D and D$^{\prime}$ bands probe phonons with general $\mathbf{q}\neq 0$, and several diagrams need to be considered for the calculation of the matrix $\mathcal{M}=\sum_{p}\mathcal{M}_{p}$ in Eq. (\ref{eq:idr}), only a very small region of phonon phase space and a small number of diagrams contribute dominantly to the Raman intensity. In particular, numerical calculations have previously shown that the Raman cross section is mostly due to a very small region in phonon phase space associated with the backscattering of the resonant photoexcited electron-hole pair\cite{thomsen2004} [see Fig. \ref{fig:schematics}(a)] and, additionally, dominated by the $ab$ diagrams\cite{venezuela2011} in the right column of Fig. \ref{fig:diagrams} [see Fig. \ref{fig:schematics}(c)]. These two results were explained in terms of so-called phase interference effects.\cite{thomsen2004,venezuela2011} In this section and in the Appendix, we quantitatively analyze these interference effects, which will allow us to significantly simplify the analytical calculation of $\mathcal{I}_{\mathrm{DR}}$ in Eq. (\ref{eq:idr}).

The fact that back-scattering of the photoexcited electron-hole pair dominates the Raman cross-section is not straight-forward to obtain only by inspection of Eqs. (\ref{eq:kee}) and (\ref{eq:keh}). A simple phase-space argument allows us to anticipate that two regions of phonon phase space are relevant, namely, $|\mathbf{q}|\sim 0$, and $|\mathbf{q}|\sim E_{\mathrm{L}}/v_{\mathrm{F}}$ [see Fig. \ref{fig:schematics}(a)-(b)]. When $|\mathbf{q}|\sim 0$, then a large number of electronic states with wave vector $|\mathbf{k}|=E_{\mathrm{L}}/2v_{\mathrm{F}}$ in Eqs. (\ref{eq:kee}) and (\ref{eq:keh}) are doubly resonant, which may lead to a proportionately large scattering amplitude. Alternatively, we note that the DR condition can only be met at some point in electronic $\mathbf{k}$ space when $|\mathbf{q}| \leq E_{\mathrm{L}}/v_{\mathrm{F}}$. Therefore, when $|\mathbf{q}|\sim E_{\mathrm{L}}/v_{\mathrm{F}}$, a singular behavior in the density of states between the photoexcited state and the back-scattered state is obtained. As shown in the Appendix, after performing the $\mathbf{k}$-variable integration in Eqs. (\ref{eq:kee}) and (\ref{eq:keh}), we obtain a significantly larger value of $|\mathcal{M}(|\mathbf{q}|\sim E_{\mathrm{L}}/v_{\mathrm{F}},\mu)|^2$ compared to $|\mathcal{M}_{p}(\mathbf{q}\rightarrow 0,\mu)|^2$ by a factor of $\omega_{\mathbf{q},\mu}^2\gamma/E_{\mathrm{L}}^3\sim 10^{-5}$.

Similarly to the G$^{\prime}$ band, $ab$ diagrams play an important role in the Raman intensity of the D and D$^{\prime}$ bands, as was first pointed out by Venezuela \textit{et at}.\cite{venezuela2011} In Appendix \ref{sec:appqbs}, we find that the poles in Eq. (\ref{eq:kee}) are differently distributed in the upper and lower complex planes from those of Eq. (\ref{eq:keh}), resulting in a matrix element $\mathcal{M}_{p}$ for $ab$ processes larger than those for $aa$ processes by a factor $\sim\omega_{\mathbf{q},\mu}/2E_{\mathrm{L}}$, as shown in Fig. \ref{fig:schematics}. Thus, failure to include $ab$ processes in the Raman calculations leads to a Raman intensity reduced by a factor $(\omega_{\mathbf{q},\mu}/2E_{\mathrm{L}})^2\sim 10^{-3}$.

A final simplification in the Raman intensity calculation is possible. As shown in Fig. \ref{fig:schematics}(c), if we consider in detail the resonance conditions in the denominators of Eq. (\ref{eq:keh}), we find that the matrix elements $\mathcal{M}_{eh}(\mathbf{q},\mu)$ is peaked exactly at $v_{\mathrm{F}}|\mathbf{q}|=E_{\mathrm{L}}$ (so-called incident light-resonance) and at $v_{\mathrm{F}}|\mathbf{q}|=E_{\mathrm{L}}-\omega_{\mathbf{q},\mu}/2$ (here both the first and third intermediate states in Eq. (\ref{eq:keh}) are at resonance with the photon). A similar conclusion holds for the $he$ diagram. Alternatively, for $\bar{eh}$ and $\bar{he}$, the peak in the matrix element occurs at $v_{\mathrm{F}}|\mathbf{q}|=E_{\mathrm{L}}-\omega_{\mathbf{q},\mu}/2$ and $v_{\mathrm{F}}|\mathbf{q}|=E_{\mathrm{L}}$ (scattered light resonance). Therefore, close to the wavevector $v_{\mathrm{F}}|\mathbf{q}|\approx E_{\mathrm{L}}$ ($v_{\mathrm{F}}|\mathbf{q}|\approx E_{\mathrm{L}}-\omega_{\mathbf{q},\mu}$), only the diagrams $\mathcal{M}_{eh}+\mathcal{M}_{he}$ ($\mathcal{M}_{\bar{eh}}+\mathcal{M}_{\bar{he}}$) need to be calculated. On the contrary, the large contribution produced by \textit{each} of the four $ab$ diagrams at $v_{\mathrm{F}}|\mathbf{q}|=E_{\mathrm{L}}-\omega_{\mathbf{q},\mu}/2$ interfere destructively, as shown in Fig. \ref{fig:schematics} and discussed in the Appendix \ref{sec:appqbs}.

\subsection{Integrated Raman Intensity}

The two-peak shape of the Raman scattering matrix shown in Fig. \ref{fig:schematics} and originating from the diagrams $eh+he$ at $v_{\mathrm{F}}|\mathbf{q}|=E_{\mathrm{L}}$, and from $\bar{eh}+\bar{he}$ at $v_{\mathrm{F}}|\mathbf{q}|=E_{\mathrm{L}}-\omega_{\mathbf{q},\mu}$, significantly simplifies the calculation of the integrated Raman intensity, as it is now only necessary to study $\mathcal{M}(\mathbf{q},\mu)$ in the close vicinity of these peaks. For this purpose, we calculate $\mathcal{M}_{eh}(\mathbf{q},\mu)+\mathcal{M}_{he}(\mathbf{q},\mu)$ for wavevectors $|\mathbf{q}|=E_{\mathrm{L}}/v_{\mathrm{F}} + \delta q$, with $|\delta q|\ll \omega_{\mathbf{q},\mu}/v_{\mathrm{F}}$. Calculation of $\mathcal{M}_{\bar{eh}}(\mathbf{q},\mu)+\mathcal{M}_{\bar{he}}(\mathbf{q},\mu)$ can be done analogously. As shown in Appendix \ref{sec:appqbs}, we find  that $\mathcal{M}_{eh}(\mathbf{q},\mu)$ is given by
\begin{equation}
\mathcal{M}_{eh}(\mathbf{q},\mu) =- \frac{iA\mathcal{K}_{\mathbf{q},\mu}}{8 v_{\mathrm{F}}^{2}\omega_{\mathbf{q},\mu}}\sqrt{\frac{2 E_{\mathrm{L}}}{ (v_{\mathrm{F}}|\mathbf{q}|-E_{\mathrm{L}}) + i\gamma/2}},
\label{eq:keh1}
\end{equation}
where $\mathcal{K}_{\mathbf{q},\mu}$ is the product of the four matrix elements in the numerator of Eq. (\ref{eq:keh}) with initial wave vector $\mathbf{k}=-\mathbf{q}/2$, so that the electron-phonon interaction couples electronic states with momentum $\mathbf{q/2}$ and $-\mathbf{q}/2$. Specifically, the value of $\mathcal{K}_{\mathbf{q},\mu}$ is given by
\begin{equation}
\mathcal{K}_{\mathbf{q},\mu}=\sum_{s,j}\frac{2\pi (ev_{\mathrm{F}})^2F_\mu \mathcal{U}_{\mathbf{q},\mu}^{e}[ \mathbf{e}_{\mathbf{Q}_{\mathrm{i}}\lambda_{\mathrm{i}}}\times\hat{\mathbf{q}}]_z[ \mathbf{e}_{\mathbf{Q}_{\mathrm{f}}\lambda_{\mathrm{f}}}^*\times\hat{\mathbf{q}}]_z e^{-i\mathbf{q}\cdot\mathbf{r}_j}}{V E_{\mathrm{L}}  \sqrt{A^3\rho\omega_{\mathbf{q},\mu}}},
\label{eq:matel}
\end{equation}
where the term $\mathcal{U}_{\mathbf{q},\mu}^{e}$ is the short-hand notation for the matrix element $\mathcal{U}_{\mathbf{q},\mu}^{e}=\langle \mathbf{q}/2,\pi^*,s'| U_{\mathbf{q}}e^{i\mathbf{q}\cdot\mathbf{r}}| -\mathbf{q}/2,\pi^{*},K \rangle $, with $s'=K$ for $\mu=\Gamma$ [i.e., projects on the diagonal component $U_{\mathbf{q},\Gamma}$ in Eq. (\ref{eq:Hdc})], and $s'=K^{\prime}$ for $\mu=K$ [i.e., projects on the off-diagonal component $U_{\mathbf{q},K}$ in Eq. (\ref{eq:Hdc})]. Importantly, in Eq. (\ref{eq:matel}), both valleys contribute to $\mathcal{M}_{eh}(\mathbf{q},\Gamma)$ for intravalley scattering, whereas only one valley contributes to $\mathcal{M}_{p}(\mathbf{q},K)$ for intervalley scattering (the creation of a phonon at the $K$ point allows an electronic transition from the $K^{\prime}$ to the $K$ point, but not vice versa). A similar analysis can be done for $\mathcal{M}_{he}(\mathbf{q},\mu)$, where hole scattering by the defects yields a matrix element $\mathcal{U}_{\mathbf{q},\mu}^{h}=\langle \mathbf{q}/2,\pi,s'| U_{\mathbf{q}}e^{i\mathbf{q}\cdot\mathbf{r}}| -\mathbf{q}/2,\pi,K \rangle $, where $s'=K$ for $\mu=\Gamma$, and $s'=K^{\prime}$ for $\mu=K$, and resulting in a total defect scattering matrix element $\mathcal{U}_{\mathbf{q},\mu}=\mathcal{U}_{\mathbf{q},\mu}^{e}-\mathcal{U}_{\mathbf{q},\mu}^h$.\cite{vtot}

In order to obtain the integrated Raman intensity, we sum $\mathcal{M}_{eh}(\mathbf{q},\mu)$ and $\mathcal{M}_{he}(\mathbf{q},\mu)$ and insert the sum in Eq. (\ref{eq:idr2}). In the regime of uncorrelated defects, $\sum_{j,j'}e^{i\mathbf{q}\cdot(\mathbf{r}_{j}-\mathbf{r}_{j'})}/A=n_{\mathrm{i}}$, where $n_{\mathrm{i}}$ is the defect concentration. Furthermore, because of the isotropic nature of the Dirac Hamiltonian, we can assume that $|\mathcal{U}_{\mathbf{q},\mu}|^2$ depends only on the modulus of the wave vector $\mathbf{q}$. Integration  over all possible phonon momenta and photon polarization directions, and considering detection of the backscattered photons, leads to the dimensionless Raman intensity
\begin{equation}
\frac{d\mathcal{I}_{\mathrm{DR}}^{\mu}}{d\Omega_{\mathrm{f}}} = \frac{g_{\mu}\alpha^{2}}{4} \frac{F_{\mu}^2}{\rho v_{\mathrm{F}}^2\omega_{\mathbf{q},\mu}}\left(\frac{v_{\mathrm{F}}}{c} \frac{E_{\mathrm{L}}}{\omega_{\mathbf{q},\mu}} \right )^{2} \frac{n_{\mathrm{i}}|\mathcal{U}_{\mathbf{q},\mu}|^{2}}{v_{\mathrm{F}}^{2}}
\mathrm{ln}\left( \frac{\omega_{\mathbf{q},\mu}}{\gamma} \right),
\label{eq:Idr}
\end{equation}
for the D ($\mu=K$) and D$^{\prime}$ ($\mu=\Gamma$) Raman process, where $\alpha=e^{2}/ c$ is the fine-structure constant, $|\mathbf{q}|=E_{\mathrm{L}}/v_{\mathrm{F}}$, and the prefactors $g_{\Gamma}=2$ and $g_{K}=1$ appear due to the different electron and phonon valley indices summations for intra and intervalley processes, respectively (see details in Appendix \ref{sec:appqbs}).

\subsection{Comparison with experiments}
\label{sec:exp}

Several experiments measured the Raman intensity ratio $I_{\mathrm{D}}/I_{\mathrm{G}}$ as a function of laser energy.\cite{pimentadefect,cancadodefect,barros2007,eckmann2013} The dependence of $\mathcal{I}_{\mathrm{DR}}$ on $E_{\mathrm{L}}$ in Eq. (\ref{eq:Idr}) is affected by several factors: (i) the resonant electronic and phonon phase space increases at larger values of photon energies; (ii) because of the dispersive behavior of the D and D$^{\prime}$ bands, $\omega_{\mathbf{q},\mu}$ varies as the laser energy is changed; (iii) the broadening $\gamma$ depends on the energy of the resonant photoexcited electron hole pairs and, in the simplest case, $\gamma$ behaves as $\gamma\propto E_{\mathrm{L}}$;\cite{elifetime} (iv) the Raman process selects specific Fourier components $|\mathcal{U}_{\mathbf{q},\mu}|^2$ of the scattering potential, with $|\mathbf{q}|=E_{\mathrm{L}}/v_{\mathrm{F}}$. Although (i) and (ii) are factors associated with the intrinsic properties of graphene, (iii) and (iv) are extrinsic and explain why different dependencies of the D-band intensity on laser energy are measured experimentally. 

Considering a linear dependence of the inverse electronic lifetime with laser energy, and the dispersion relation of the $A_1$ phonon mode close to the $K$ point, we plot in Fig. \ref{fig:exp} the intensity ratio $I_{\mathrm{D}}/I_{\mathrm{G}}$ as a function of $E_{\mathrm{L}}$ for point-like defects (i.e., $|\mathcal{U}_{\mathbf{q},\mu}|^2$ is taken as independent of $\mathbf{q}$). The analytical results are compared with the experimental integrated Raman intensity from Ref. \onlinecite{cancadodefect}. For the $I_{\mathrm{G}}$ Raman intensity, we use the standard textbook dependence $I_{\mathrm{G}}\propto E_{\mathrm{L}}^4$,\cite{basko2008} and we used typical values for the electronic broadening $\gamma \sim 0.03 E_{\mathrm{L}}$.\cite{elifetime} Even within the simplifying assumptions made in our model, there is good agreement between theory and the experiments.

Furthermore, it is interesting to note that, from Eq. (\ref{eq:Idr}), the disorder-induced D and D$^{\prime}$-band intensities do not necessarily have the same dependence on $E_{\mathrm{L}}$. In fact, recent experimental measurements\cite{eckmann2013} have shown that the ratio $I_{\mathrm{D}^{\prime}}/I_{\mathrm{D}}$ is a slowly increasing function of laser energy. If we consider point-like defects and taking into account that $\gamma\ll\omega_{\mathbf{q},\mu}$, then the ratio $I_{\mathrm{D}^{\prime}}/I_{\mathrm{D}}$ obtained from Eq. (\ref{eq:Idr}) verifies  $I_{\mathrm{D}^{\prime}}/I_{\mathrm{D}}\propto (\omega_{\mathbf{q},K}/\omega_{\mathbf{q},\Gamma})^3$, where $|\mathbf{q}|=E_{\mathrm{L}}/v_{\mathrm{F}}$. Because the $A_1$ phonon mode near the $K$ point is more dispersive than the LO phonon mode near the $\Gamma$ point, then the ratio $I_{\mathrm{D}^{\prime}}/I_{\mathrm{D}}$ obtained from theory is a slowly increasing function of laser energy, which is in agreement with the experiments.

We finally consider the dependence of the integrated Raman intensity on defect concentration $n_{\mathrm{i}}$. Within the model in Eq. (\ref{eq:Idr}), two regimes exist: (i) when the defect concentration $n_i$ is low enough such that the electron-phonon induced linewidth $\gamma_{\mathrm{ep}}\sim$15 meV\cite{elifetime} is larger than the defect induced linewidth $\gamma_{\mathrm{ed}}$, then $I_{\mathrm{D}}\propto n_{\mathrm{i}}$; (ii) however, when $n_{\mathrm{i}}$ is sufficiently large such that $\gamma_{\mathrm{ed}} > \gamma_{\mathrm{ep}}$, then $\gamma$ is sensitive to defect concentration $n_{\mathrm{i}}$ and a non-linear dependence of $\mathcal{I}_{\mathrm{DR}}$ as a function of $n_{\mathrm{i}}$ is obtained. The threshold value of $n_{\mathrm{i}}$ separating both regimes can be estimated by calculating the defect-induced broadening of the electronic states at $\varepsilon_{\mathbf{k}}\sim E_{\mathrm{L}}/2$, assuming uncorrelated short-range defects with a potential strength $|\mathcal{U}_{\mathbf{q},\mu}|=U_0$. A straight-forward calculation yields $\gamma_{\mathrm{ed}}=n_{\mathrm{i}}|U_0|^2 E_{\mathrm{L}}/2(\hbar v_{\mathrm{F}})^2$. Taking $U_0 \sim 1$ eV$\cdot$nm$^{2}$ and $E_{\mathrm{L}}\sim 2$ eV, then the condition $\gamma_{\mathrm{ep}}\sim\gamma_{\mathrm{ed}}$ is met at defect concentrations of $n_{\mathrm{i}}\sim 10^{12}$ cm$^{-2}$.

In order to compare with experimental measurements, the dependence of $I_{\mathrm{D}}$ on $n_{\mathrm{i}}$ is plotted  in Fig. \ref{fig:exp}(b) together with the experimental data from Ref. \onlinecite{concentration2}. Here we used $\gamma_{\mathrm{ep}}\sim15$ meV and $\gamma_{\mathrm{ed}}[\mathrm{meV}]\sim 10\times n_{\mathrm{i}}[10^{12}\mathrm{cm}^{-2}]$. The theoretical model correctly captures the saturating behavior of the D-band intensity, as obtained in experiments. However, it is beyond the scope of this paper to describe the highly defective limit, such as that measured in Refs. \onlinecite{pimentadefect,cancadodefect}. In this limit, the electronic states are localized within small grains formed, for instance, after intense ion irradiation and thus, they can no longer be described as eigenstates of the translational invariant system.

\section{Discussion}
\label{sec:discussion}

The defect scattering potential plays an important role in determining the DR Raman intensity, as shown in Eq. (\ref{eq:Idr}). However, most models to date typically assume constant elastic scattering matrix elements. First, this is equivalent to assuming that defects can scatter electrons or holes with equal strength throughout the BZ. Second, this assumption neglects electronic phase factors associated with the sublattice and valley pseudospin degrees of freedom. For instance, whether the on-site component of the defect potential provides a significantly different contribution to the Raman intensity than the hopping component has not been addressed in the literature. Thus, further work on the analysis of the term $|\mathcal{U}_{\mathbf{q},\mu}|^2$, which conveniently appears as a numerical prefactor in Eq. (\ref{eq:Idr}), is necessary.

Experimental measurements for different types of defects have shown $I_{\mathrm{D}}\gg I_{\mathrm{D}^{\prime}}$.\cite{eckmann2012} By taking the ratio of Eq. (\ref{eq:Idr}) for the D and D$^{\prime}$ bands, we obtain
\begin{equation}
\frac{I_{\mathrm{D}}}{I_{\mathrm{D}^{\prime}}}\approx \frac{g_{K}}{g_{\Gamma}}\frac{F_{K}^2}{F_{\Gamma}^2} \left(\frac{\omega_{\mathbf{q},\Gamma}}{\omega_{\mathbf{q},K}}\right)^{3} \frac{|\mathcal{U}_{\mathbf{q},K}|^{2}}{|\mathcal{U}_{\mathbf{q},\Gamma}|^{2}}\approx 2.2\times \frac{|\mathcal{U}_{\mathbf{q},K}|^{2}}{|\mathcal{U}_{\mathbf{q},\Gamma}|^{2}}.
\label{eq:ratio}
\end{equation}
Although theoretical calculations show $F_{\Gamma}<F_{K}$ (or more precisely, $F_{\Gamma}/F_{K}\approx 1/\sqrt{2}$),\cite{lazzeridft,jiangeph,andointer} this small difference cannot account for the large intensity ratio observed experimentally. Additionally, the phonon frequencies verify $\omega_{\mathbf{q},\Gamma}/\omega_{\mathbf{q},K} \approx 1.3 $. Then, Eq. (\ref{eq:ratio}) suggests that the origin of $I_{\mathrm{D}}/I_{\mathrm{D}^\prime}\gg 1$ is primarily due to the scattering potential term.

The fact that short-wave-vector intravalley scattering typically dominates over long-wave-vector intervalley scattering suggests that there is a contradiction between Eq. (\ref{eq:ratio}) and the typically measured relation $I_{\mathrm{D}}/I_{\mathrm{D}^{\prime}}\gg 1$. In particular, when the defect potential has a finite range, the short-wave-vector scattering components of the matrix $U_{\mathbf{q},\Gamma}$ in Eq. (\ref{eq:Hdc}) are expected to be larger than the long-wave-vector scattering components in $U_{\mathbf{q},K}$. However, this does not necessarily mean $|\mathcal{U}_{\mathbf{q},K}|<|\mathcal{U}_{\mathbf{q},\Gamma}|$. Because graphene has internal pseudospin degrees of freedom, the internal phases of the photoexcited electron (or hole) and the backscattered electron (or hole) play an important role. In particular, it is well-known from the behavior of the electronic transport of graphene that intravalley backscattering of Dirac electrons is strongly suppressed,\cite{andobackscatt1,andobackscatt2} thereby allowing $|\mathcal{U}_{\mathbf{q},K}|>|\mathcal{U}_{\mathbf{q},\Gamma}|$ to be possible. Similar effects are expected to occur for the DR theory, where backscattering of the photoexcited electrons [see Fig. \ref{fig:schematics}(a)] is the dominant contribution to the DR Raman intensity. Further theoretical work in this direction is necessary and should be the subject of future studies.

Using Raman spectroscopy to identify the nature of the defects may have attractive applications in the characterization of real graphene samples. For instance, it has been previously found\cite{baskoedge} that the edge-induced D-band intensity scales with laser energy as $I_{\mathrm{D}}\propto E_{\mathrm{L}}\mathrm{ln}(\omega_{\mathbf{q},K}/\gamma)$, which is significantly different from the dependence found in Eq. (\ref{eq:Idr}). Therefore, our result suggests a way to distinguish the edge-induced D band from the disorder-induced D band. Alternatively, defects with different ranges may be distinguished between each other by the different wave-vector dependence of the term $|\mathcal{U}_{\mathbf{q},\mu}|^2$. In practice, however, extracting such information may be difficult given that several parameters in Eq. (\ref{eq:Idr}) change simultaneously with laser energy, thus making detailed experimental analysis rather complicated. It is more likely, however, that use of the ratio $I_{\mathrm{D}}/I_{\mathrm{D}^{\prime}}$ is a more promising direction to identify the nature of defects, as suggested by Eckmann \textit{et al}.\cite{eckmann2012}

\section{Conclusions}
\label{sec:conclusions}

A detailed analytical study of the disorder-induced double resonant (DR) Raman process in graphene was presented, and analytical expressions for the Raman probability $\mathcal{I}_{\mathrm{DR}}$ for the D and D$^{\prime}$ bands are derived and discussed. Given the large number of parameters required to describe the DR process, this study succeeds in explicitly showing how the Raman intensities depends on laser energy, defect concentration, and electronic lifetime, within a single equation [Eq. (\ref{eq:Idr})]. Furthermore, we here discussed quantitatively the so-called phase interference effects,\cite{thomsen2004,venezuela2011} which determine the most important phonon wave vectors and diagrams in Fig. \ref{fig:diagrams} that contribute to the DR Raman intensity. It was also found that the disorder-induced D-band Raman intensity has a different laser energy dependence than the edge-induced D band,\cite{baskoedge} which could potentially be used to distinguish carrier scattering by boundaries from scattering due to lattice disorder.

Good agreement between our analytical results and experimental measurements is obtained. As observed experimentally, it is shown in this paper that the D- and D$^{\prime}$-band intensities have a different laser energy dependence\cite{eckmann2013} and, additionally, that each of these dependencies can vary with the type of defect.\cite{pimentadefect,cancadodefect,barros2007,eckmann2013} The saturating behavior of the $I_{\mathrm{D}}$ intensity with increasing defect concentrations measured in experiments\cite{concentration,concentration2} is also discussed, and occurs when the defect collision rate is faster than the electron-phonon collision rate. Further theoretical work is required to better understand the role of the different parameters describing the defect scattering potential, such as the range and the various components associated with the electronic pseudo-spin degrees of freedom, on determining the $I_{\mathrm{D}}/I_{\mathrm{D}^{\prime}}$ ratio. The value of this ratio could potentially be used to identify the nature of defects in graphene.\cite{eckmann2012}

\begin{acknowledgments}
We thank A. Jorio, L. G. Cancado and E. H. Martins Ferreira for providing experimental data. JFRN and MSD acknowledge support from Grant NSF/DMR1004147. EBB acknowledges support from CNPq Grant No. 245640/2012-6 and FUNCAP. RS acknowledges MEXT grant Nos. 25107005 and 25286005.
\end{acknowledgments}


\bibliography{dband.bib}

\begin{thebibliography}{10}

\bibitem{milliebook}
A.~Jorio, R.~Saito, G.~Dresselhaus, and M.~S. Dresselhaus, {\it Raman
  Spectroscopy in Graphene Related Systems} (Wiley-VCH, Berlin, 2011).

\bibitem{saitoreview}
R.~Saito, M.~Hofmann, G.~Dresselhaus, A.~Jorio, and M.~S. Dresselhaus, Advances
  in Physics {\bf 60}(3), 413--550 (2011).

\bibitem{ferrarireview}
A.~C. Ferrari and D.~M. Basko, Nat Nano {\bf 8}(4), 235 (2013).

\bibitem{phonondisp}
A.~Gr\"uneis, R.~Saito, T.~Kimura, L.~G. Can\ifmmode~\mbox{\c{c}}\else
  \c{c}\fi{}ado, M.~A. Pimenta, A.~Jorio, A.~G. Souza~Filho, G.~Dresselhaus,
  and M.~S. Dresselhaus, Phys. Rev. B {\bf 65}, 155405 (2002).

\bibitem{ferrarireview2}
A.~C. Ferrari, Solid State Communications {\bf 143}(1–2), 47 -- 57 (2007).
\newblock Exploring graphene Recent research advances.

\bibitem{malardreview}
L.M. Malard, M.A. Pimenta, G.~Dresselhaus, and M.S. Dresselhaus, Physics
  Reports {\bf 473}(5–6), 51 -- 87 (2009).

\bibitem{doping1}
S.~Pisana, M.~Lazzeri, C.~Casiraghi, K.~S. Novoselov, A.~K. Geim, A.~C.
  Ferrari, and F.~Mauri, Nat Mater {\bf 6}(3), 198--201 (2007).

\bibitem{doping2}
J.~Yan, Y.~Zhang, P.~Kim, and A.~Pinczuk, Phys. Rev. Lett. {\bf 98}, 166802
  (2007).

\bibitem{gphlayers2006}
A.~C. Ferrari, J.~C. Meyer, V.~Scardaci, C.~Casiraghi, M.~Lazzeri, F.~Mauri,
  S.~Piscanec, D.~Jiang, K.~S. Novoselov, S.~Roth, and A.~K. Geim, Phys. Rev.
  Lett. {\bf 97}, 187401 (2006).

\bibitem{edges1}
L.~G. Can\ifmmode~\mbox{\c{c}}\else \c{c}\fi{}ado, M.~A. Pimenta, B.~R.~A.
  Neves, M.~S.~S. Dantas, and A.~Jorio, Phys. Rev. Lett. {\bf 93}, 247401
  (2004).

\bibitem{edges2}
A.~K. Gupta, T.~J. Russin, H.~R. Guti\'{e}rrez, and P.~C. Eklund, ACS Nano {\bf
  3}(1), 45--52 (2009).

\bibitem{edges3}
C.~Casiraghi, A.~Hartschuh, H.~Qian, S.~Piscanec, C.~Georgi, A.~Fasoli, K.~S.
  Novoselov, D.~M. Basko, and A.~C. Ferrari, Nano Letters {\bf 9}(4),
  1433--1441 (2009).
\newblock PMID: 19290608.

\bibitem{pimentadefect}
M.~A. Pimenta, G.~Dresselhaus, M.~S. Dresselhaus, L.~G. Cancado, A.~Jorio, and
  R.~Saito, Phys. Chem. Chem. Phys. {\bf 9}, 1276--1290 (2007).

\bibitem{cancadodefect}
L.~G. Cançado, A.~Jorio, E.~H.~Martins Ferreira, F.~Stavale, C.~A. Achete,
  R.~B. Capaz, M.~V.~O. Moutinho, A.~Lombardo, T.~S. Kulmala, and A.~C.
  Ferrari, Nano Letters {\bf 11}(8), 3190--3196 (2011).

\bibitem{drthomsen}
C.~Thomsen and S.~Reich, Phys. Rev. Lett. {\bf 85}, 5214--5217 (2000).

\bibitem{c887}
R.~Saito, A.~Jorio, A.~G. {Souza~Filho}, G.~Dresselhaus, M.~S. Dresselhaus, and
  M.~A. Pimenta, Phys. Rev. Lett. {\bf 88}, 027401 (2001).

\bibitem{f1020}
R.~Saito, A.~Gr{\"u}neis, {Ge.}~G. Samsonidze, V.~W. Brar, G.~Dresselhaus,
  M.~S. Dresselhaus, A.~Jorio, L.~G. Can\c{c}ado, C.~Fantini, M.~A. Pimenta,
  and A.~G. {Souza~Filho}, New Journal of Physics {\bf 5} (2003).

\bibitem{thomsen2004}
J.~Maultzsch, S.~Reich, and C.~Thomsen, Phys. Rev. B {\bf 70}, 155403 (2004).

\bibitem{reichreview}
S.~Reich and C.~Thomsen, Philosophical Transactions of the Royal Society of
  London. Series A: Mathematical, Physical and Engineering Sciences {\bf
  362}(1824), 2271--2288 (2004).

\bibitem{ramanreview}
Mildred~S. Dresselhaus, Ado Jorio, Mario Hofmann, Gene Dresselhaus, and
  Riichiro Saito, Nano Letters {\bf 10}(3), 751--758 (2010).

\bibitem{dispersive1}
R.P. Vidano, D.B. Fischbach, L.J. Willis, and T.M. Loehr, Solid State
  Communications {\bf 39}(2), 341 -- 344 (1981).

\bibitem{dispersive2}
Y.~Wang, D.~C. Alsmeyer, and R.~L. McCreery, Chemistry of Materials {\bf 2}(5),
  557--563 (1990).

\bibitem{basko2008}
D.~M. Basko, Phys. Rev. B {\bf 78}, 125418 (2008).

\bibitem{concentration2}
E.~H. Martins~Ferreira, Marcus V.~O. Moutinho, F.~Stavale, M.~M. Lucchese,
  R.~B. Capaz, C.~A. Achete, and A.~Jorio, Phys. Rev. B {\bf 82}, 125429
  (2010).

\bibitem{defectreview}
H.~Terrones, R.~Lv, M.~Terrones, and M.~S. Dresselhaus, Reports on Progress in
  Physics {\bf 75}(6), 062501 (2012).

\bibitem{barros2007}
E.~B. Barros, H.~Son, Ge.~G. Samsonidze, A.~G. Souza~Filho, J.~Mendes~Filho,
  G.~Dresselhaus, and M.~S. Dresselhaus, Phys. Rev. B {\bf 76}, 035444 (2007).

\bibitem{eckmann2013}
A.~Eckmann, A.~Felten, I.~Verzhbitskiy, R.~Davey, and C.~Casiraghi, Phys. Rev.
  B {\bf 88}, 035426 (2013).

\bibitem{eckmann2012}
A.~Eckmann, A.~Felten, A.~Mishchenko, L.~Britnell, R.~Krupke, K.~S. Novoselov,
  and C.~Casiraghi, Nano Letters {\bf 12}(8), 3925--3930 (2012).

\bibitem{concentration}
M.M. Lucchese, F.~Stavale, E.H.~Martins Ferreira, C.~Vilani, M.V.O. Moutinho,
  R.~B. Capaz, C.A. Achete, and A.~Jorio, Carbon {\bf 48}(5), 1592 -- 1597
  (2010).

\bibitem{venezuela2011}
P.~Venezuela, M.~Lazzeri, and F.~Mauri, Phys. Rev. B {\bf 84}, 035433 (2011).

\bibitem{barros2011}
E.~B. Barros, K.~Sato, Ge.~G. Samsonidze, A.~G. Souza~Filho, M.~S. Dresselhaus,
  and R.~Saito, Phys. Rev. B {\bf 83}, 245435 (2011).

\bibitem{Sato2006}
K.~Sato, R.~Saito, Y.~Oyama, J.~Jiang, L.G. Cançado, M.A. Pimenta, A.~Jorio,
  Ge.G. Samsonidze, G.~Dresselhaus, and M.S. Dresselhaus, Chemical Physics
  Letters {\bf 427}(1–3), 117 -- 121 (2006).

\bibitem{isotopes}
J.~F. Rodriguez-Nieva, R.~Saito, S.~D. Costa, and M.~S. Dresselhaus, Phys. Rev.
  B {\bf 85}, 245406 (2012).

\bibitem{basko2007}
D.~M. Basko, Phys. Rev. B {\bf 76}, 081405 (2007).

\bibitem{baskoedge}
D.~M. Basko, Phys. Rev. B {\bf 79}, 205428 (2009).

\bibitem{elifetime}
C.-H. Park, F.~Giustino, M.~L. Cohen, and S.~G. Louie, Phys. Rev. Lett. {\bf
  99}, 086804 (2007).

\bibitem{sasaki2008}
K.~Sasaki and R.~Saito, Progress of Theoretical Physics Supplement {\bf 176},
  253--278 (2008).

\bibitem{andointra}
K.~Ishikawa and T.~Ando, Journal of the Physical Society of Japan {\bf 75}(8),
  084713 (2006).

\bibitem{andointer}
H.~Suzuura and T.~Ando, Journal of the Physical Society of Japan {\bf 77}(4),
  044703 (2008).

\bibitem{andobackscatt1}
T.~Ando and T.~Nakanishi, Journal of the Physical Society of Japan {\bf 67}(5),
  1704--1713 (1998).

\bibitem{vtot}
The minus sign is due to an overall minus sign in the electron phonon matrix
  element, when hole scattering (as opposed to electron scattering) is
  considered.

\bibitem{lazzeridft}
S.~Piscanec, M.~Lazzeri, F.~Mauri, A.~C. Ferrari, and J.~Robertson, Phys. Rev.
  Lett. {\bf 93}, 185503 (2004).

\bibitem{jiangeph}
J.~Jiang, R.~Saito, Ge.~G. Samsonidze, S.~G. Chou, A.~Jorio, G.~Dresselhaus,
  and M.~S. Dresselhaus, Phys. Rev. B {\bf 72}, 235408 (2005).

\bibitem{andobackscatt2}
T.~Ando, T.~Nakanishi, and R.~Saito, Journal of the Physical Society of Japan
  {\bf 67}(8), 2857--2862 (1998).

\end{thebibliography}

\appendix

\section{Raman Intensity Calculations}
\label{sec:appcalc}

In this appendix, we focus specifically on the calculation of the $ee$ and $eh$ diagrams in Fig. \ref{fig:diagrams}. Extension to the remaining processes is straight-forward. In Sec. \ref{sec:appqbs}, we consider the most relevant case of backscattering of the photoexcited electron-hole pair due to the production of a phonon with wavevector $\mathbf{q}_{\mathrm{ph}}=\mathbf{q}_{\mu}+\mathbf{q}$ ($\mu=\Gamma$, K), where $|\mathbf{q}|\approx E_{\mathrm{L}}/v_{\mathrm{F}}$, $\mathbf{q}_{\Gamma}=0$ and $\mathbf{q}_{K}=\mathbf{K}$. Afterwards, in Sec. \ref{sec:appq0}, we show that forward-scattering of the photoexcited electron-hole pair (i.e., $|\mathbf{q}|=0$) provides a negligible contribution to the total intensity (this is shown rigorously for nanotubes in Ref. \onlinecite{thomsen2004}).

\subsection{Backscattering: $v_{\mathrm{F}}|\mathbf{q}|=E_{\mathrm{L}}$}
\label{sec:appqbs}

\begin{figure}
\centering \includegraphics[scale=1.0]{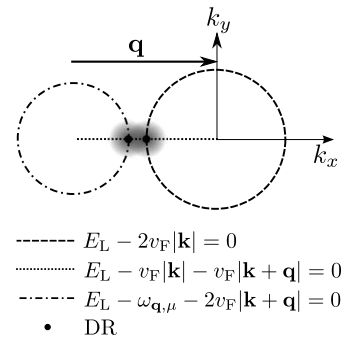}
\caption{Plot of the resonance conditions for each of the terms in the denominator of Eq. (\ref{eq:keh}), for the case $|\mathbf{q}|=E_{\mathrm{L}}/v_{\mathrm{F}}$. The shaded regions indicate the volume of electronic phase space $\mathbf{k}$ that mostly contributes to the scattering amplitude.}
\label{fig:resonance}
\end{figure}

We evaluate first the matrix element $\mathcal{M}_{p}(\mathbf{q},\mu)$ for a value of $|\mathbf{q}|=E_{\mathrm{L}}/v_{\mathrm{F}}+\delta q$, where $|\delta q| \ll \omega_{\mathbf{q},\mu}/v_{\mathrm{F}}$. Given that trigonal warping effects are neglected, we can arbitrarily align the $k_x$ direction in the integrals in Eqs. (\ref{eq:kee}) and (\ref{eq:keh}) with $\mathbf{q}$, as shown in Fig. \ref{fig:resonance}. Under the assumption $\gamma\ll\omega_{\mathbf{q},\mu}\ll E_{\mathrm{L}}$, which is the typical situation in experiments, most of the contribution to $\mathcal{M}_{p}(\mathbf{q},\mu)$ comes from the electronic phase-space region in the vicinity of the point $\mathbf{k}\approx-\mathbf{q}/2$ (shaded regions in Fig. \ref{fig:resonance}). Given the small region of phase space that needs to be considered, we: (a) expand to leading order in the vicinity of $\mathbf{k}=-\mathbf{q}/2$ the three functions in the denominators of Eqs. (\ref{eq:kee}) and (\ref{eq:keh}); (b) evaluate the matrix elements at $\mathbf{k}=-\mathbf{q}/2$; (c) perform the $\mathbf{k}$-space integration.

After carrying out the steps (a) and (b) above, and conveniently normalizing the integrals in Eqs. (\ref{eq:kee}) and (\ref{eq:keh}), one can then obtain
\begin{equation}
\mathcal{M}_{p}(\mathbf{q},\mu)\approx\frac{A\mathcal{K}_{\mathbf{q},\mu} }{8\pi^{2}v_{\mathrm{F}}^{2} E_{\mathrm{L}}} \times \mathcal{I}_{\pm}\left( \frac{v_{\mathrm{F}}\delta q}{E_{\mathrm{L}}}\right),
\label{eq:kehdim}
\end{equation}
where $\mathcal{I}_{\pm}(\xi)$ is given by
\begin{equation}
\begin{array}{ll}
\displaystyle \mathcal{I}_{\pm}(\xi)= & \displaystyle \pm\int_{-\infty}^{\infty} dx\int_{-\infty}^{\infty} dy \frac{1}{(-\xi-\omega_{\mathbf{q},\mu}/E_{\mathrm{L}} \pm x - i \epsilon)} \\
&\displaystyle \qquad\quad \times \frac{1}{(-\xi -2 y^{2}-i\epsilon)(-\xi + x - i \epsilon)},
\end{array}
\label{eq:idq}
\end{equation}
and  $\mathcal{K}_{\mathbf{q},\mu}$ is described in Eq. (\ref{eq:matel}). The $+$ ($-$) sign in Eq. (\ref{eq:idq}) corresponds to the $ee$ ($eh$) process, and $\epsilon=\gamma/2E_{\mathrm{L}}\ll 1$. 

The positions of the poles in the $x$ variable are distributed differently in the upper and lower-half planes for the $\mathcal{I}_{\pm}$ integrals, which results in $|\mathcal{I}_{+}|\ll |\mathcal{I}_{-}|$ (i.e., the dominant contribution comes from $eh$ processes). In particular, calculation of $\mathcal{I}_{-}$ in Eq. (\ref{eq:idq}) yields
\begin{equation}
\mathcal{I}_{-}(\xi)=-\frac{i\pi^2E_{\mathrm{L}}}{\omega_{\mathbf{q},\mu}}\sqrt{\frac{2}{\xi+i\epsilon}}.
\label{eq:kehdq}
\end{equation}
On the other hand, for the $ee$ process, $\mathcal{I}_+=0$ is obtained when using the approximations discussed above. However, the leading-order correction to $\mathcal{I}_+$ can be estimated to be order $\mathcal{I}_+(0)/\mathcal{I}_-(0) \sim -i (\omega_{\mathbf{q},\mu}/2E_{\mathrm{L}})$, which is consistent with the numerical results in Fig. \ref{fig:schematics}. Therefore, the $aa$ diagrams lead to a substantially smaller scattering amplitude $\sim(\omega_{\mathbf{q},\mu}/2E_{\mathrm{L}})^2$ when compared to the $ab$ diagrams, and this feature was previously pointed out in the work by Venezuela \textit{et al}.\cite{venezuela2011} Inserting $\mathcal{I}_{-}$ into Eq. (\ref{eq:kehdim}) yields Eq. (\ref{eq:keh1}).

At $v_{\mathrm{F}}|\mathbf{q}|=E_{\mathrm{L}}$, the $he$ process also contributes strongly to the Raman intensity, while all remaining $ab$ processes provide a small contribution ($\bar{eh}+\bar{he}$ are peaked at $v_{\mathrm{F}}|\mathbf{q}|=E_{\mathrm{L}}-\omega_{\mathbf{q},\mu}$). In order to calculate the integrated Raman intensity, we insert $\mathcal{M}_{eh}(\mathbf{q},\mu)+\mathcal{M}_{he}(\mathbf{q},\mu)$ into Eq. (\ref{eq:idr2}) to obtain
\begin{equation}
\begin{array}{ll}
\displaystyle\frac{d\mathcal{I}_{\mathrm{DR}}^{\mu}}{d\Omega_{\mathrm{f}}} = & \displaystyle \frac{g_{\mu}\alpha^2}{16}\frac{F_{\mu}^2}{\rho v_{\mathrm{F}}^2 \omega_{\mathbf{q},\mu}}\frac{v_{\mathrm{F}}^2}{c^2} \sum_{\mathbf{q},\lambda_{\mathrm{f}}}\frac{n_i|\mathcal{U}_{\mathbf{q},\mu}|^2}{A\omega_{\mathbf{q},\mu}^2} \\ 
&  \displaystyle \qquad \times \frac{E_{\mathrm{L}}|\mathbf{e}_{\mathbf{Q}_{\mathrm{i}}\lambda_{\mathrm{i}}}\times\hat{\mathbf{q}}|^2|\mathbf{e}_{\mathbf{Q}_{\mathrm{f}}\lambda_{\mathrm{f}}}^*\times\hat{\mathbf{q}}|^2}{\sqrt{(v_{\mathrm{F}}|\mathbf{q}|-E_{\mathrm{L}})^2+(\gamma/2)^2}}.
\end{array}
\label{eq:didr1}
\end{equation}
Here, $\alpha=e^{2}/ c$ is the fine-structure constant, $g_{\Gamma}=2$, $g_{K}=1$, and where we used the assumption of uncorrelated defects with a concentration $n_i$. Different prefactors $g_{\mu}$ appear for intravalley and intervalley processes because, for zone-center phonons, both valleys contribute to $\mathcal{K}_{\mathbf{q},\mu}$, as discussed in the main text, while for zone-boundary phonons, only one valley contributes to \textit{each} phonon mode in the vicinity of the $K$ and $K^{\prime}$ points. 

Integration over momentum space $\mathbf{q}$ in Eq. (\ref{eq:didr1}) can be done in the vicinity of a ring of radius $E_{\mathrm{L}}/v_{\mathrm{F}}$ and angular direction $\theta_{\mathbf{q}}$. Thus, we use polar coordinates $\sum_{\mathbf{q}}\approx (A/2\pi)\int d (\delta q) \int d\theta_{\mathbf{q}}(E_{\mathrm{L}}/v_{\mathrm{F}})$. Furthermore, we assume normal and unpolarized incident photons, and  detection in both polarization directions. Then, the angular integration of Eq. (\ref{eq:didr1}) yields
\begin{equation}
\sum_{\lambda_{\mathrm{f}}}\int \frac{d\theta_{\mathbf{q}}}{2\pi} |\mathbf{e}_{\mathbf{Q}_{\mathrm{i}}\lambda_{\mathrm{i}}}\times\hat{\mathbf{q}}|^2|\mathbf{e}_{\mathbf{Q}_{\mathrm{f}}\lambda_{\mathrm{f}}}^*\times\hat{\mathbf{q}}|^2=\frac{1+\mathrm{cos}^2\theta_{\mathrm{f}}}{2},
\label{eq:didr2}
\end{equation}
where $\theta_{\mathrm{f}}$ is the angle of the outgoing photon with respect to the normal to the graphene sheet. Detection in the backscattering configuration (i.e., $\theta_{\mathrm{f}}=\pi$) is assumed in this work. The radial integration of Eq. (\ref{eq:didr1}), using a cutoff in the phonon momentum of $\sim \omega_{\mathbf{q},\mu}/2v_{\mathrm{F}}$, which is the region of validity of Eq. (\ref{eq:kehdq}) (see Fig. \ref{fig:schematics}), yields half the value of the integrated Raman intensity of Eq. (\ref{eq:Idr}). The other half of the value of the integrated Raman intensity comes from considering the peak at $v_{\mathrm{F}} |\mathbf{q}|=E_{\mathrm{L}}-\omega_{\mathbf{q},\mu}$ from the $\bar{eh}+\bar{he}$ diagrams.

We finally note that the peak at $v_{\mathrm{F}}|\mathbf{q}|=E_{\mathrm{L}}-\omega_{\mathbf{q},\mu}/2$ provides a negligible contribution to $\sum_{p}\mathcal{M}_{p}$, as shown in Fig. \ref{fig:schematics}. In this case, the large contribution of $\mathcal{M}_{eh}$ cancels that of $\mathcal{M}_{\bar{eh}}$ when each term is calculated separately as in Eqs. (\ref{eq:kehdim}) and (\ref{eq:idq}). Similarly, the contribution $\mathcal{M}_{he}$ cancels that of $\mathcal{M}_{\bar{he}}$, yielding a negligible value of $\mathcal{M}=\sum_{p}\mathcal{M}_{p}$. 

\subsection{Forward-Scattering: $\mathbf{q}=0$}
\label{sec:appq0}

Forward-scattering [Fig. \ref{fig:schematics}(b)] provides a negligible contribution to the D and D$^{\prime}$-band intensities because of the small scattering amplitude when compared to those associated with the backward scattering case, $v_{\mathrm{F}}|\mathbf{q}|=E_{\mathrm{L}}$. To show this point, we compute the matrix element $\mathcal{M}_{eh}(\mathbf{q}\rightarrow 0,\Gamma)$ for the zone-center phonon mode, which is given by
\begin{widetext}
\begin{equation}
\mathcal{M}_{eh}(\mathbf{q}\rightarrow 0,\Gamma) = A \int_{0}^{\infty} dk \, k\int_0^{2\pi}\frac{d\theta_{\mathbf{k}}}{2\pi} \frac{\mathcal{K}_{\mathbf{q}\rightarrow 0,\Gamma}(\theta_{\mathbf{k}})}{(E_{\mathrm{L}}-\omega_{\mathbf{q},\mu}-2v_{\mathrm{F}}k-i\gamma/2)(E_{\mathrm{L}}-2v_{\mathrm{F}}k-i\gamma/2)^2},
\label{eq:mq01}
\end{equation}
and where $\mathcal{K}_{\mathbf{q}\rightarrow 0,\Gamma}(\theta_{\mathbf{k}})$ is
\begin{equation}
\displaystyle \mathcal{K}_{\mathbf{q}\rightarrow 0,\Gamma}(\theta_{\mathbf{k}}) =-\sum_{s,j} \frac{2\pi(ev_{\mathrm{F}})^2F_{\Gamma} \, \mathcal{U}_{\mathbf{q}=0,\Gamma}\, \mathrm{sin}(\theta_{\mathbf{k}})\, [\mathbf{e}_{\mathbf{Q}_{\mathrm{i}}\lambda_{\mathrm{i}}}\times\hat{\boldsymbol\theta}]_z[ \mathbf{e}_{\mathbf{Q}_{\mathrm{f}}\lambda_{\mathrm{f}}}^*\times\hat{\boldsymbol\theta}]_z e^{-i\mathbf{q}\cdot\mathbf{r}_j}}{ V E_{\mathrm{L}}  \sqrt{A^3\rho\,\omega_{\mathbf{q},\mu}}}.
\label{eq:mq0}
\end{equation}
\end{widetext}
In Eqs. (\ref{eq:mq01}) and (\ref{eq:mq0}), $\theta_{\mathbf{k}}$ was chosen to be the angle beween the $\mathbf{k}$-vector and the atomic displacement $\mathbf{u}$, and $\hat{\boldsymbol\theta}=[\mathrm{cos}(\theta_{\mathbf{k}}),\mathrm{sin}(\theta_{\mathbf{k}})]$. Integration of the radial and angular components of Eq.\,(\ref{eq:mq0}) yields
\begin{equation}
\mathcal{M}_{eh}(\mathbf{q}\rightarrow 0,\Gamma)=\frac{A \langle  \mathcal{K}_{\mathbf{q}\rightarrow 0,\Gamma} \rangle_{\theta} \mathcal{L}_{k}}{4 E_{\mathrm{L}} v_{\mathrm{F}}^{2}}, 
\label{eq:mq02}
\end{equation}
where $\langle  \mathcal{K}_{\mathbf{q}\rightarrow 0,\Gamma} \rangle_{\theta}=\int (d\theta_{\mathbf{k}}/2\pi)\mathcal{K}_{\mathbf{q}\rightarrow 0,\Gamma}(\theta_{\mathbf{k}})$, and $\mathcal{L}_{k}$ is
\begin{equation}
\mathcal{L}_{k}=\frac{E_{\mathrm{L}}}{\omega_{\mathbf{q},\mu}}\left[1-\frac{E_{\mathrm{L}}-\omega_{\mathbf{q},\mu}-i\frac{\gamma}{2}}{\omega_{\mathbf{q},\mu}}\mathrm{ln}\left( \frac{E_{\mathrm{L}}-i\frac{\gamma}{2}}{E_{\mathrm{L}}-\omega_{\mathbf{q},\mu}-i\frac{\gamma}{2}}\right)\right].
\end{equation}
Considering the case $\gamma\ll\omega_{\mathbf{q},\mu}\ll E_{\mathrm{L}}$, then $\mathcal{L}_{k}\approx 1$. By comparing Eq. (\ref{eq:mq02}) with Eq. (\ref{eq:kehdq}), we conclude that $|\mathcal{M}_{eh}(\mathbf{q}\rightarrow 0,\Gamma)|^{2}$ is a factor of order $\omega_{\mathbf{q},\mu}^2\gamma/E_{\mathrm{L}}^3\sim 10^{-5}$ smaller than $|\mathcal{M}_{eh}(|\mathbf{q}|=E_{\mathrm{L}}/v_{\mathrm{F}},\Gamma)|^2$ at backscattering, for typical values $\gamma \sim 10$ meV.

\end{document}